\RequirePackage[colorlinks,allcolors=blue,breaklinks=true]{hyperref}
\documentclass[final]{agujournal2019}
\usepackage{apacite}
\usepackage{lmodern}
\usepackage{mathrsfs}
\usepackage{amsmath}
\usepackage{amssymb}
\usepackage{amsfonts}
\usepackage{bm}
\usepackage[english]{babel}
\usepackage{url} 
\newcommand*{\doi}{}

\makeatletter
\newcommand{\doi@}[1]{\urlstyle{same}\url{https://doi.org/#1}}
\DeclareRobustCommand{\doi}{\hyper@normalise\doi@}
\makeatother
\usepackage{units}
\usepackage{color}
\usepackage{calc}
\usepackage{graphicx}
\usepackage{gensymb}
\usepackage{wasysym}
\usepackage[capitalize,nameinlink]{cleveref}
\usepackage{lineno}
\usepackage{soul}
\usepackage[normalem]{ulem}
\journalname{JGR: Space Physics}

\begin{document}
%
%

\title{Electron-scale energy transfer due to lower hybrid waves during asymmetric reconnection}

%
%

\authors{Sabrina F. Tigik\affil{1}, Daniel B. Graham\affil{1}, Yuri V. Khotyaintsev\affil{1}}
\affiliation{1}{Swedish Institute of Space Physics, Uppsala, Sweden}



\correspondingauthor{Sabrina F. Tigik and Daniel B. Graham}{sabrina.tigik@irfu.se, dgraham@irfu.se}

\begin{keypoints}
  \item Spectral analysis is used to quantify energy exchange between lower hybrid waves and electrons during magnetopause reconnection.
  \item The energy exchange occurs within a thin plasma mixing layer at the magnetospheric outflow edge, and is highly fluctuating.
  \item On average, the electrons gain energy from waves, possibly contributing to the observed electron heating. 
\end{keypoints}

%
%

%
%

\begin{abstract}
  We use Magnetospheric Multiscale (MMS) mission data to investigate electron-scale energy transfer due to lower hybrid drift waves during magnetopause reconnection. We analyze waves observed in an electron-scale plasma mixing layer at the edge of the magnetospheric outflow. Using high-resolution $7.5~\unit{ms}$ electron moments, we obtain an electron current density with a Nyquist frequency of $\sim 66~\unit{Hz}$, sufficient to resolve most of the lower hybrid drift wave power observed in the event. We then employ wavelet analysis to evaluate $\delta{\bf J\, \cdot}\,\delta{\bf E}$, which accounts for the phase differences between the fluctuating quantities. The analysis shows that the energy exchange is localized within the plasma mixing layer, and it is highly fluctuating, with energy bouncing between waves and electrons throughout the analyzed time and frequency range. However, the cumulative sum over time indicates a net energy transfer from the waves to electrons. We observe an anomalous electron flow toward the magnetosphere, consistent with diffusion and electron mixing. These results indicate that waves and electrons interact dynamically to dissipate the excess internal energy accumulated by sharp density gradients. We conclude that the electron temperature profile within the plasma mixing layer is produced by a combination of electron diffusion across the layer, as well as heating by large-scale parallel potential and lower hybrid drift waves.
\end{abstract}


\section{Introduction}
\label{sec:intro}
Magnetic reconnection is a relaxation process in which rapid change in field topology releases energy accumulated in the magnetic fields to the charged particles, leading to plasma heating and acceleration. Exactly how energy injected by large-scale structures is dissipated at kinetic scales in rarefied plasmas, where Coulomb collisions are unimportant, is an unsolved fundamental problem. A prominent hypothesis considers the effects of plasma waves excited by kinetic instabilities, which are ubiquitous in collisionless magnetic reconnection \cite{khotyaintsev2019}, acting as a proxy for energy transfer, leading to particle acceleration (\emph{anomalous} transport/diffusion) \cite{davidson1975,gary1980,gurnett1983,treumann1991,pavan2013} and energy dissipation (\emph{anomalous} heating/resistivity) \cite{gary1971,drake2003,yoon2006b,che2017,khotyaintsev2020}. In this scenario, the energy exchange between kinetic waves and particles could promote an efficient relaxation of unstable phase-space configurations introduced by changes in the magnetic field topology, resulting in acceleration, diffusion, and heating of the plasma particles.

At Earth's low-latitude magnetopause, magnetic reconnection occurs between the shocked solar wind plasma of the magnetosheath and the magnetospheric plasma. Due to the distinct properties of the reconnecting plasmas, where the magnetosheath plasma typically has a weaker magnetic field and considerably higher density than the magnetospheric plasma, magnetopause reconnection is usually asymmetric. Unlike the symmetric case, in asymmetric reconnection, the X-line and the stagnation point are decoupled \cite{ugai2000,swisdak2003}, with the stagnation point shifted towards the low-density side of the X-line. Such a displacement allows for a net flow of plasma through the X-line from the high-density side towards the low-density side of the reconnecting boundary \cite{cassak2007}. In magnetopause reconnection, this means that the denser and less energetic magnetosheath plasma crosses the magnetopause and mixes with the highly energetic but extremely tenuous magnetosphere plasma in the magnetospheric separatrix layer \cite{sckopke1981,hesse2017}, where large electron temperature anisotropy with $T_{e\parallel}/T_{e\perp}>1$ is commonly observed \cite{hall1991,graham2017}. The increase in $T_{e\parallel}$ is attributed to the trapping and heating of electrons by a large-scale parallel potential \cite{egedal2011,egedal2013}. However, the plasma mixing signatures and electron parallel heating at the magnetospheric separatrix region are consistently observed in the presence of intense wave activity in the lower hybrid frequency range \cite{gurnett1979,cattell1995,bale2002,khotyaintsev2016,graham2019}.

As the denser magnetosheath plasma inflow approaches the stagnation point at the low-density side of the magnetopause boundary, steep density gradients form, making the magnetospheric separatrix region unstable to the lower hybrid drift instability. Within the inhomogeneous interface between the reconnecting plasmas, the lower hybrid drift instability is triggered as a response to cross-field currents associated with density and magnetic field gradients in the plasma \cite{krall1971,yoon2004a}. Due to the nature of their source instability, lower hybrid drift waves have been suggested as a possible source of cross-field diffusion and electron mixing at the magnetopause \cite{treumann1991,vaivads2004,graham2017,le2017,price2020}. \citeA{khotyaintsev2016} reported MMS observations of an asymmetric reconnection layer, which is significantly thicker than the predictions from 2D simulations, and attributed this difference to diffusion by strong lower hybrid drift waves observed at the boundary. Recently, using high-resolution fields and particle measurements from the four MMS spacecraft \cite{burch2016}, \citeA{graham2022} reported the first direct observations of anomalous effects associated with lower hybrid drift waves in the magnetopause. Using data from different magnetopause crossings to calculate the anomalous terms for electric fields in the frequency range of lower hybrid drift waves, the authors showed that: (1) the effects of anomalous resistivity and viscosity balance each other, meaning the waves do not significantly contribute to the reconnection electric field; (2) the waves do produce significant anomalous electron diffusion across the magnetopause, possibly contributing to the relaxation of the density gradients at the magnetospheric separatrix \cite{graham2022}. 

While there is strong evidence of the significant contribution of lower hybrid drift waves to particle transport and mixing in the magnetospheric separatrix region, their contribution to electron heating in that region is still a matter of debate. Spacecraft observations \cite{graham2014b,graham2016c} have shown that most of the properties of the parallel electron heating observed in the magnetospheric separatrix region at close distances to the X-line are consistent with the large-scale parallel potential model in \citeA{egedal2011,egedal2013}. However, later studies using MMS data reported that far from the X-line, within thin sub-layers of the magnetospheric separatrix region where intense lower hybrid drift wave activity is commonly observed, $T_{\parallel e}$ decreases while the electron number density $n_e$ increases \cite{graham2017,graham2019,wang2017b,holmes2019}. The density-temperature profile in those observations is incompatible with the Boltzmann and the Chew-Goldberger-Low (CGL) scaling laws used to close the fluid equations hierarchy in the large-scale parallel potential model \cite{le2009,egedal2013}; such deviation is attributed to the enhanced lower hybrid drift wave-driven cross-field diffusion of low-energy magnetosheath electrons into the magnetospheric separatrix region, increasing the density and lowering the electron temperature in that region.

Using fully kinetic 3D simulations with parameters typical of magnetopause reconnection, \citeA{le2018} reported that lower hybrid waves driven by diamagnetic drifts along the steep density gradient in the magnetospheric separatrix region led to cross-field transport, allowing for magnetosheath electrons to cross the magnetopause and mix with the magnetospheric population in regions away from the X-line. According to the authors, the flow of cross-field diffused low-energy magnetosheath electrons into the magnetospheric separatrix region accounts for the observed decrease in the $T_{\parallel e}$ as $n_e$ increases, leading to a density-temperature profile in agreement with the profiles reported in spacecraft observations \cite{graham2017,graham2019,wang2017b,holmes2019}.  However, it is unclear if the incoming low-energy electron population gains energy from the lower hybrid drift waves during the diffusion process. In the present work, we investigate the energy exchange between lower hybrid drift waves and electrons in a thin plasma mixing layer at the edge of the magnetospheric outflow (separatrix region). We characterize the energy exchange between lower hybrid drift waves and electrons and present evidence that electrons gain energy in the process.

\section{Observations}
\label{sec:observ}
\subsection{Data}
\label{sec:data}
We use MMS's high-resolution fields and particle data to investigate signatures of energy exchange between electrons and lower hybrid waves during a magnetopause crossing recorded in burst mode on November 06, 2016, from 08:39:34 UT to 08:41:44 UT. During this interval, the MMS constellation is in a tetrahedral formation, located close to the subsolar point, towards the dusk flank, at $\sim [8.0,\, 5.0,\, -0.7]~\unit{R_E}$ in geocentric solar ecliptic (GSE) coordinates, where $R_E$ is Earth's radius. The average spacecraft separation is $\sim 11.6~\unit{km}$. We utilize electric field data sampled at $8192~\unit{Hz}$ by the electric field double probe (EDP) instrument \cite{lindqvist2016,ergun2016}, background magnetic field measured by the flux-gate magnetometer (FGM) \cite{russell2016} at a sample rate of $128~\unit{Hz}$, and wave magnetic field from the search-coil magnetometer (SCM) \cite{lecontel2016} sampled at $8192~\unit{Hz}$. For the analysis, we transform all the vector data to the boundary normal (LMN) coordinate system by applying the minimum variance analysis method to the magnetic field ${\bf B}$ measured in GSE coordinates by MMS1 between 08:40:52.5 and 08:41:07.3. The resulting transformation vectors from GSE to LMN coordinates are ${\bf \hat L}=(0.40,\, -0.01,\, 0.91)$, ${\bf \hat M} = (0.52,\, -0.82,\, -0.24)$, and ${\bf \hat N}=(0.75,\, 0.57,\, -0.32)$, where ${\bf \hat L}$ has the maximum variance in ${\bf B}$ and is directed along the reconnecting component of ${\bf B}$; ${\bf \hat N}$, the minimum ${\bf B}$ variance component, is normal to the magnetopause boundary and points towards the Sun; ${\bf \hat M} = {\bf \hat N \times \hat L}$ completes the unit vector triad lying along the X-line direction. 

For the particles, we start with burst data from the fast plasma investigation (FPI) instrument \cite{pollock2016} sampled every $30~\unit{ms}$ for electrons and $150~\unit{ms}$ for ions. Then, similar to the method described in \citeA{rager2018}, we use the time offset data for each of the individual azimuthal deflections of FPI/DES to obtain particle distributions at a higher time resolution, $7.5~\unit{ms}$ ($133~\unit{Hz}$) for electrons and $37.5~\unit{ms}$ ($27~\unit{Hz}$) for protons. These higher sample rates are achieved by reducing the azimuthal angular resolution of the particle distributions. However, such a compromise in angular resolution does not significantly affect the present analysis since we are not investigating fine structures in the particles' velocity distributions but rather the bulk particle properties, i.e., moments. With a Nyquist frequency of $\sim 66~\unit{Hz}$, the electron current density ${\bf J}$ calculated using the high-resolution electron moments can resolve the frequency range comprising most of the lower hybrid drift wave power observed in the event under analysis, enabling an effective ${\bf J \cdot E}$ investigation at electron scales. \Cref{sec:method} presents a detailed description of the methodology we use in this analysis.

\subsection{Event overview}
\label{sec:event}
\begin{figure}[h!]
  \noindent \centering \includegraphics[width=0.99\textwidth]{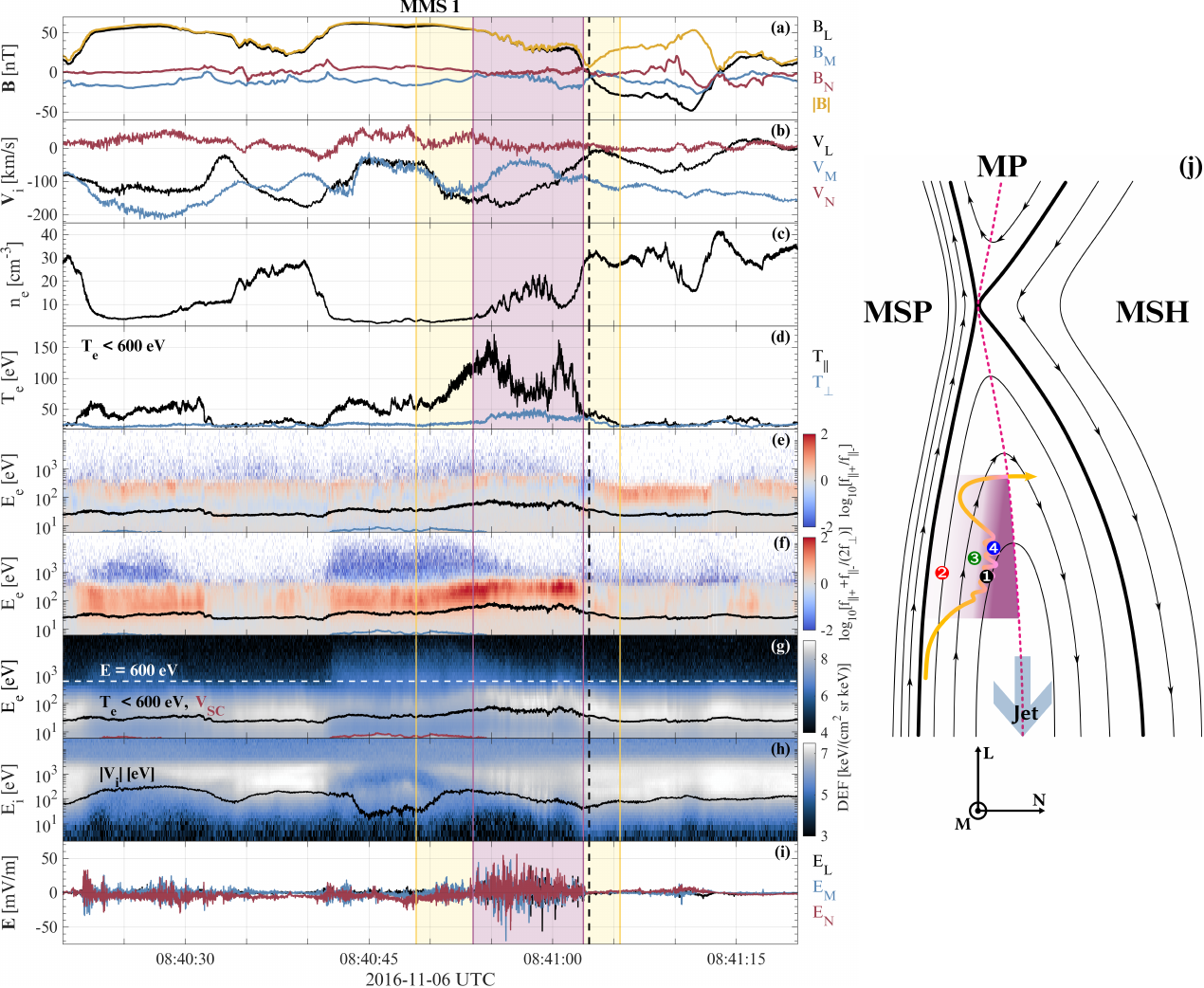}
  \caption{Overview plot of the event as measured by MMS1. Vector quantities are in LMN coordinates. (a) Magnetic field vector and absolute value. (b) Ion velocity vector. (c) Electron number density. (d) Parallel and perpendicular electron temperatures. (e) The ratio between parallel and anti-parallel electron phase-space density. (f) The ratio between field-aligned and perpendicular electron phase-space density. (g) Omnidirectional electron differential energy flux. The white dashed line delimits the maximum cutoff energy of $600~\unit{eV}$ used to exclude highly energetic magnetospheric population when calculating the electrons temperature, given by the black line. The red line is the spacecraft's potential, $V_{SC}$. (h) Omnidirectional ion differential energy flux. The black line indicates the energy associated with ${\bf V}_i$. (i) Electric field vector. The region in yellow marks the four spacecraft's trajectory represented by the yellow path in the sketch. (j) Idealized sketch of the magnetopause crossing in the ${\bf N}$--${\bf L}$ plane. The region in dark magenta marks the interval in which the four MMS travel through the thin plasma mixing layer at the border of the magnetospheric outflow.}
  \label{fig:1}
\end{figure}

Figures \ref{fig:1}a-\ref{fig:1}i show a one-minute overview, from 08:40:20 UTC to 08:41:20 UTC, of the magnetopause crossing under investigation, as measured by MMS1. At the beginning of the interval, \Cref{fig:1}a shows a rapid increase in ${\bf B}$, indicating that the spacecraft is moving towards the magnetosphere after a partial crossing (omitted). From $\sim$ 08:40:32 to $\sim$ 08:40:42, ${\bf B}$ decreases and becomes more turbulent, and the electron number density in \Cref{fig:1}c increases to magnetosheath levels. In \Cref{fig:1}b, the reconnection jet reaching $V_{iL}\approx -180~\unit{km/s}$ by 08:40:40, indicating an underway nearby reconnection \cite{retino2005}. However, the spacecraft does not cross the null point ($B_L=0$) before gradually retreating toward the magnetosphere, indicating another partial crossing. The cutoff in the ${\bf -\hat M}$ ion flow, from 08:40:44 to 08:40:48, indicates that energetic magnetosheath ions subjected to the finite gyroradius effect cannot reach this region, implying that during this interval the spacecraft is beyond the ion edge \cite{oieroset2015,phan2016b,graham2017}, going even further into the inner magnetospheric separatrix region than the previous retreat towards the magnetospheric side. However, the electron number density $n_e\gtrsim 2$ in \Cref{fig:1}c, and the presence of a considerable flux of low energy electrons with $E_e< 600~\unit{eV}$ in the differential electron energy flux in \Cref{fig:1}g, indicates that MMS does not cross the electron edge towards the magnetosphere proper and remains inside the magnetospheric separatrix region.

Similar to the ion edge, the electron edge is the maximum distance that the most energetic magnetosheath electrons can reach into the magnetospheric separatrix region. Because of their different parallel speeds, ions and electrons sharing the same transverse drift have distinct penetration boundaries, with the electrons usually reaching much further into the magnetospheric separatrix region than the ions, a phenomenon known as \emph{time-of-flight} effect \cite{gosling1990}.  The electron edge is typically located by the inner side of the magnetospheric separatrix and is generally a reliable indication of the first open magnetic field line. Closer to the magnetopause, the ion edge usually coincides with the reconnection jet/outflow boundary \cite{gosling1990,lindstedt2009,oieroset2015}. These edges mark important transition regions within the magnetospheric separatrix region \textemdash a small-scale segment of the low-latitude boundary layer close to the X-line at the subsolar magnetopause, with the same structure and plasma mixing properties as the large-scale boundary. Within the magnetospheric separatrix region, as the spacecraft moves towards the magnetopause, the tenuous and energetic magnetospheric-like plasma close to the electron edge becomes increasingly denser after the ion edge, reaching magnetosheath-like densities in the outflow region \cite{sckopke1981,hall1991}. In \Cref{fig:1}c, from 08:40:40 until the vertical dashed line marking the null point crossing, the electron number density variations are consistent with the spacecraft transiting within the magnetospheric separatrix region.

Given the partial crossings and the gradual transitions between magnetospheric-like and magnetosheath-like environments, we estimate that the spacecraft formation approaches the magnetopause in a nearly parallel trajectory to the boundary. The final and complete crossing in this event occurs from 08:40:48.85 to 08:41:05.05\textemdash the yellow interval in the overview figure. We define the beginning of the magnetopause crossing as the point where the spacecraft crosses the ion edge towards the magnetopause, seen in \Cref{fig:1}b as a progressive increase in the ${\bf -\hat M}$ ion flow, which also appears as a progressive increase in the flux of magnetosheath ions in \Cref{fig:1}h. In addition, \Cref{fig:1}b shows a reconnection jet in the ${\bf -\hat L}$ direction, indicating that the spacecraft is approaching the magnetopause south of the X-line. The ion jet lasts for approximately $14~\unit{s}$, from $\sim$ 08:40:49 to $\sim$ 08:41:03, and reaches $V_{iL}\approx -180~\unit{km/s}$ around 08:40:56. The absence of a reconnection jet reversal indicates that the spacecraft does not cross the X-line in this event. In \Cref{fig:1}a, ${\bf B}$ shows a slow and eventful magnetopause crossing attached to a flux rope. The vertical dashed line marks the time in which MMS1 crossed $B_L=0$ at 08:41:03. A few moments later, the yellow region ends at 08:41:05.50, when \Cref{fig:1}d shows $T_{e\parallel}/ T_{e\perp}\approx1$, indicating the magnetopause crossing is complete and the spacecraft is in the magnetosheath.

\Cref{fig:1}j shows an idealized sketch of the full magnetopause crossing highlighted in the overview figure, along with the relative spacecraft separations in the ${\bf N}$--${\bf L}$ plane. The yellow arrow shows an estimate of the MMS formation's trajectory within the interval highlighted in yellow in the overview figure. We focus this analysis on the nine-second interval highlighted in dark magenta in the data panels (a-j) and in the schematic (panel j) where we observe large-amplitude oscillations, with frequencies $\gtrsim 1~\unit{Hz}$, in ${\bf B}$ and $n_e$, and intense electric field fluctuations in \Cref{fig:1}i. The electric field fluctuations have frequencies comparable to the local lower hybrid frequency, which we identify as lower hybrid drift waves (discussed below). The oscillations occur while the MMS constellation travels along a thin plasma mixing layer at the edge of the magnetospheric outflow, and due to the back-and-forth movement of the thin boundary layer through the MMS formation. The wavy path within the dark magenta region (panel j) represents the oscillatory boundary motion along the MMS path, while the dark magenta color gradient gives a simplified depiction of the expected plasma density variation along ${\bf \hat N}$.

In \Cref{fig:1}a, inside the dark magenta region, ${\bf |B|}$ decreases from $50~\unit{nT}$ to $35~\unit{nT}$ while we observe intense $\gtrsim 1~\unit{Hz}$ oscillations first in $B_L$, and then in $B_M$ ($\sim$ 08:40:56.5). The oscillations in $B_M$ last until $\sim$ 08:41:00, when we observe a sharp increase in $B_M$ from  $|B_M| \approx 10~\unit{nT}$ to $|B_M| \approx 20~\unit{nT}$. From 08:41:00.5 to 08:41:02.5, $B_M$ accounts for most of the sudden increase observed in ${\bf |B|}$ since $B_L$ and $B_N$ remain nearly constant. After 08:41:02.5, $B_M$ decreases quickly, returning to values close to zero around  08:41:03.5. A remarkable feature of the magnetic field in this event is that, unlike most of the magnetic reconnection events in the magnetopause, $B_L$ has nearly the same magnitude on both sides of the magnetopause boundary. This strong $B_L$ in the magnetosheath side is related to the large flux rope observed by MMS after it crossed the magnetopause.

In \Cref{fig:1}d, we show the parallel and perpendicular temperatures for electrons with energies $< 600~\unit{eV}$. This energy cutoff is necessary to exclude suprathermal magnetospheric electrons. As MMS moves towards the magnetopause boundary, $T_{e\parallel}$ increases from $\sim 50~\unit{eV}$ at 08:40:50 to $\sim 120~\unit{eV}$ at 08:40:53.5\textemdash the beginning of the dark magenta region. Inside the dark magenta interval, $T_{e\parallel}$ has two maximum regions of $\sim 170~\unit{eV}$, one around 08:40:54.5 and another around 08:41:01.5. Both maxima are located by the limits of the dark magenta interval. At $\sim$ 08:40:55.5, $T_{e\parallel}$ starts to decrease at a similar rate as the increase in $n_e$, seen in \Cref{fig:1}c. In the center of the dark magenta interval, from 08:40:56.5 to 08:40:59.5, $T_{e\parallel}$ reaches its lowest values, fluctuating around $50\lesssim T_{e\parallel}\lesssim 100 ~\unit{eV}$, while $n_e$ oscillates between $12 \lesssim n_e \lesssim 25~\unit{cm^{-3}}$. The significant increase in $n_e$ is possibly related to the cross-field transport of magnetosheath electrons through the magnetopause boundary driven by the intense lower hybrid drift waves observed within the same interval. As discussed in the introduction, this enhanced flux of low-energy electrons from the magnetosheath forms a plasma mixing layer, which leads to the observed decrease in $T_{e\parallel}$. As for the perpendicular electrons, we also observe a sharp increase in $T_{e\perp}$ around 08:40:54.5, which remains fluctuating around $T_{e\perp} \approx 45~\unit{eV}$ at the center of the dark magenta interval.

Figures \ref{fig:1}e and \ref{fig:1}f show how the electron energy is distributed in the directions parallel, anti-parallel, and perpendicular to ${\bf B}$. \Cref{fig:1}e shows the ratio between parallel and anti-parallel electron phase-space densities, $f_{\parallel+}/f_{\parallel-}$. Gray shaded regions mark the time and energy range where $f_{\parallel+}\approx f_{\parallel-}$, red hued regions mark $f_{\parallel+}>f_{\parallel-}$, and blue hued regions display $f_{\parallel+}<f_{\parallel-}$. Through most of the figure, we observe a predominantly gray shade at energies $< 600~\unit{eV}$, which indicates that $f_{\parallel+} $ and $f_{\parallel-}$ are mostly balanced. The only relevant exception is the region surrounding $B_L=0$, where we see a parallel jet at $E\lesssim 100~\unit{eV}$, and an anti-parallel jet at $100\lesssim E \lesssim 600 ~\unit{eV}$. Similarly, \Cref{fig:1}f shows the ratio between the total field-aligned electron phase-space density and the perpendicular electron phase-space density, $(f_{\parallel+}+f_{\parallel-})/2f_{\perp}$. At suprathermal energies ($E>600~\unit{eV}$), the gradual decrease in the perpendicular phase-space density anisotropy associated with energetic magnetospheric electrons indicates that the spacecraft is slowly approaching the magnetopause boundary. At $E<600~\unit{eV}$, the red colored regions indicate temperature anisotropy with $T_{e\parallel}/T_{e\perp}>1$. As expected, the largest temperature anisotropy regions are collocated with the two $T_{e\parallel}$ peaks in \Cref{fig:1}d. From $\sim$08:40:55 to $\sim$08:41:00, the parallel anisotropy is lower and highly variable, with the highest values seen at energies within $150\lesssim E \lesssim 500 ~\unit{eV}$.

From the description above, it is clear that the electron energization in this event is not trivial. For instance, the parallel energization observed at the center of the dark magenta region has different properties from those observed in both extremities of the interval, where the peaks in $T_{e\parallel}$ occur. The omnidirectional electron differential energy flux in \Cref{fig:1}g is rather uniform around thermal energy levels during the intense electron parallel energization observed in both borders of the dark magenta region. In these two regions, we also observe flattop electron velocity distributions (not shown) at pitch-angles $\theta = 0^{\degree} $ and $180^{\degree} $, and rather Maxwellian electrons at $\theta = 90^{\degree}$. These characteristics are consistent with heating by large-scale parallel potential \cite{egedal2011,egedal2013}. However, as the spacecraft approaches the center of the dark magenta region (after $\sim$ 08:40:55), the oscillatory pattern observed $n_e$ is seen in \Cref{fig:1}f as variations in the spectrogram intensity at energies $70\lesssim E \lesssim 200 ~\unit{eV}$. This means that the electrons crossing the magnetopause boundary, producing the density increase and the formation of the thin plasma mixing layer observed during this interval, have energies well above the average magnetosheath temperature $T_{e\rm{MSH}}\approx 22 ~\unit{eV}$. The higher energy levels observed in the thin plasma mixing layer raise the question of whether the lower hybrid drift waves contribute to the observed electron heating.

\section{Plasma mixing layer}
\label{sec:bound-layer-thickn}
\begin{figure}[h!]
  \centering \includegraphics[width=0.99\linewidth]{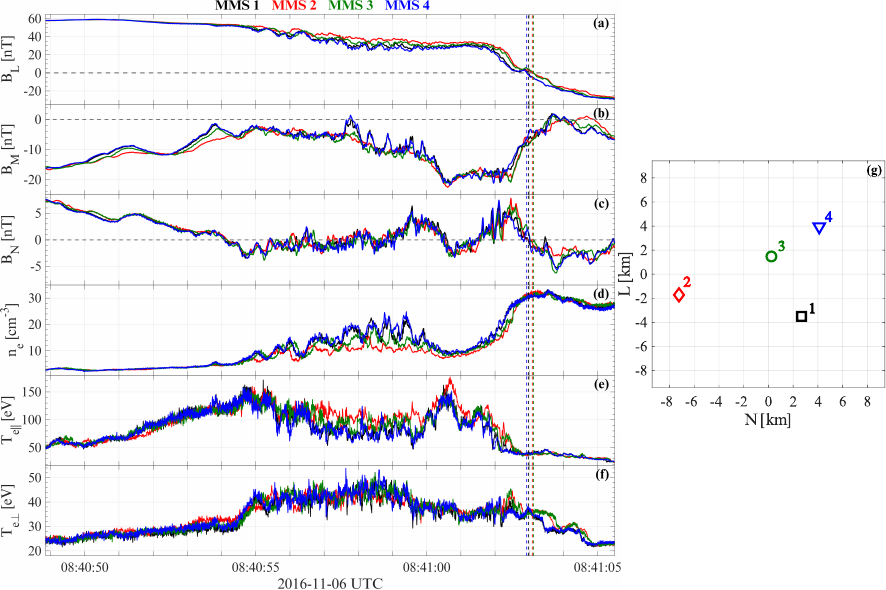}
  \caption{Measurements obtained by each spacecraft. The lines are color-coded following the legend on the top of the figure. (a-f) $B_L$, $B_M$, $B_N$, $n_e$, $T_{e\parallel}$, and $T_{e\perp}$. The gray dashed line in (a) marks $B_L=0$, and the vertical dashed lines show the moment each spacecraft crosses $B_L=0$. (g) Projection of the MMS formation on the ${\bf N}$--${\bf L}$ plane.}
  \label{fig:2}
\end{figure}

In this section, we use multi-spacecraft analysis to estimate the thickness of the plasma mixing layer observed in the dark magenta interval in \Cref{fig:1}. Figures \ref{fig:2}a-\ref{fig:2}f show four spacecraft measurements of $B_L$, $B_M$, $B_N$, $n_e$, $T_{e\parallel}$, and $T_{e\perp}$, respectively, in the interval marked in yellow in \Cref{fig:1}. \Cref{fig:2}g shows a projection of the MMS formation in the ${\bf N}$--${\bf L}$ plane, with $+L$ indicating northward displacement from the tetrahedron center, and $+N$ indicating Sunward displacement from the tetrahedron center.

To estimate the plasma mixing layer thickness, we consider the relative spacecraft separation in the ${\bf \hat N}$ direction, and differences in the measurements of $B_L$, $B_M$, $n_e$ and $T_{e\parallel}$ made by each spacecraft within the time interval marked in dark magenta in \Cref{fig:1}. The other two quantities, $B_N$ and $T_{e\perp}$, are shown for the sake of completeness since both show negligible measurement differences among the four spacecraft. In \Cref{fig:2}c, $B_N$ is roughly the same for the four spacecraft, with all of them showing a small bipolar enhancement of $\sim 7~\unit{nT}$ around the null point, which might indicate a nearby flux transfer event \cite{paschmann1982,owen2001}. Similarly, \Cref{fig:2}f shows the same $T_{e\perp}$ features and levels for all the spacecraft. However, both $n_e$ and $T_{e||}$ show significant differences between the spacecraft; the largest difference is between MMS4 and MMS2, which have the largest separation in the N-direction.

We start the analysis with MMS4, which is the farthest spacecraft towards the magnetosheath. Located at $R_N \approx 4.25~\unit{km}~{\bf \hat N}$, MMS4 observes the highest amplitude oscillations in $B_L$, $B_M$, and $n_e$ from $\sim$ 08:40:54.5 to $\sim$ 08:41:00.0. We attribute these oscillations to the movement of the boundary layer through the spacecraft, where the $n_e$ depletions correspond to the boundary moving towards the magnetosheath, leaving the spacecraft farther inside magnetospheric separatrix region. In the same way, the density enhancements correspond to the boundary moving earthwards, leaving the spacecraft inside the magnetospheric outflow region. Thus, it is reasonable to assume that MMS4 is well inside the plasma mixing layer, close to the outflow boundary. Based on this assumption and the MMS4 position shown in figure \Cref{fig:2}g, we define MMS4 as the reference point of the outermost border of the plasma mixing layer to estimate the boundary thickness.

MMS1 is separated from MMS4 by $\sim -1.5~\unit{km}~{\bf \hat N}$, which in magnitude is slightly below the electron inertial length averaged over the four spacecraft, $\langle d_e\rangle \approx 1.7~\unit{km}$. Both spacecraft have very similar measurements in Figures \ref{fig:2}a-\ref{fig:2}f, suggesting the plasma mixing layer properties are relatively constant at distances $\lesssim \langle d_e\rangle$. MMS3 is $\sim -4~\unit{km}~{\bf \hat N}$ ($\sim 2.3\langle d_e\rangle$) away from MMS4, and $\sim -2.5~\unit{km}~{\bf \hat N}$ ($\sim 1.4\langle d_e\rangle$) from MMS1. Despite the very small separation from MMS1, MMS3 observes a boundary layer with slightly different properties, showing irregular oscillations with lower amplitudes in $B_L$ and $n_e$, and no oscillations in $B_M$. In addition, in comparison to MMS1/4 and MMS2, MMS3 observes intermediate values of $B_L$, $n_e$, and $T_{e\parallel}$. The differences between MMS3 and MMS1/4 measurements become even more evident from $\sim$ 08:40:57.5 to $\sim$ 08:41:00.5, when the increase in density reduces the average electron inertial length to $\langle d_e\rangle \lesssim 1.5~\unit{km}$. This suggests that the spatial properties of the plasma mixing layer change gradually with the local electron inertial length.

MMS2 is separated by $\sim -11.5~\unit{km}~{\bf \hat N}$ ($\sim 6.7\langle d_e\rangle$) from MMS4. MMS2 measurements diverge from the other three spacecraft throughout the plasma mixing layer interval, remaining in a more magnetospheric-like environment where it observes higher $B_L$, lower $n_e$, and higher $T_{e\parallel}$. MMS2 also does not capture the oscillations in $B_L$, $B_M$ and $n_e$ observed by MMS1, MMS4 and MMS3 (partially). These differences are consistent with its position in \Cref{fig:2}g, which is farther inside the magnetospheric separatrix region. After 08:41:02.5, the four spacecraft show similar measurements again and cross the magnetopause a few milliseconds later in the order shown by the vertical dashed lines. This implies that MMS2 never fully entered the thin boundary layer at the edge of the magnetospheric outflow and crossed the magnetopause through a relatively more regular environment in comparison with the other spacecraft. Therefore, it is reasonable to consider the relative position between MMS4 and MMS2, with an error margin of $\sim \pm \langle d_e\rangle$, as a good estimate for the thickness of the boundary layer crossed by the MMS formation during this event, which gives us $L_b\approx 11.5~\unit{km}~\pm 1.7~\unit{km}$, or $L_b\approx 6.7\langle d_e\rangle\pm \langle d_e\rangle$.

\section{Methodology}
\label{sec:method}
To quantify the energy exchange between electrons and the lower hybrid drift waves observed within the thin plasma mixing layer, we calculate the dissipation term from Poynting's theorem, ${\bf J \cdot E}$. The objective is to analyze \emph{local} energy transfer between electric field fluctuations and electron motion fluctuations. Thus, we maintain the fluctuating electric field $\delta{\bf E}$ in the observation frame instead of switching to the electron bulk motion frame (${\bf E'} = {\bf E} + {\bf V}_e{\bf \times B}$), commonly used in studies of ${\bf J \cdot E}$ that do not focus on local interaction between fluctuating quantities but in entire electron scale structures containing dissipation regions \cite{zenitani2011}.

\begin{figure}[b!]
\centering  \includegraphics[width=0.98\linewidth]{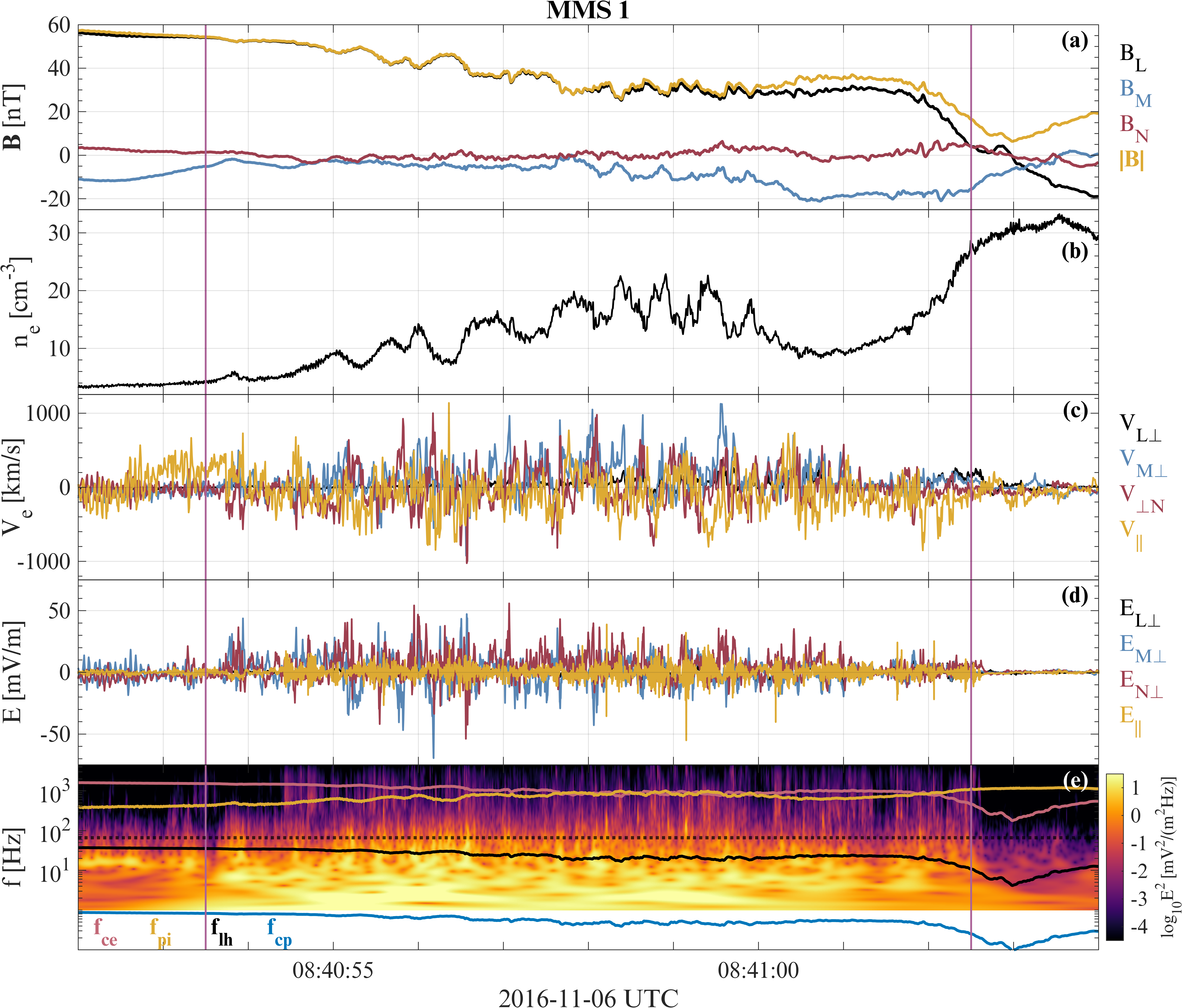}
\caption{Electron velocity and electric field within the thin boundary layer at the magnetospheric separatrix (the interval between the dark magenta lines). (a) Magnetic field vector and absolute value. (b) Electron number density. (c) Electron velocity in LMN coordinates decomposed in perpendicular and parallel components. (d) Electric field in LMN coordinates decomposed in perpendicular and parallel components. (e) Full electric field spectrum overlaid with $f_{ce}$ (red line), $f_{lh}$ (solid black line), $f_{cp}$ (blue line), and $f=66~\unit{Hz}$ (black dotted line). The white region below $1~\unit{Hz}$ in (e) is shown to include the proton cyclotron frequency line.}
\label{fig:3}
\end{figure}

We denote the fluctuating energy conversion term as $\delta{\bf J}\cdot \delta{\bf E}$. The total fluctuating current density is given by $\delta{\bf J}=-en(\delta{\bf V}_e-\delta{\bf V}_i)$, where $-e$ is the electron charge, $n$ is the plasma number density, $\delta{\bf V}_e$ is the fluctuating electron velocity, and $\delta{\bf V}_i$ is the fluctuating ion velocity. However, the only particle data capable of resolving current density fluctuations within the lower hybrid drift waves frequency range is the high-resolution electron moments, with a $133~\unit{Hz}$ sample rate. In contrast, with a sample rate of $27~\unit{Hz}$, the highest resolution ion moments can resolve only frequencies below $13.3~\unit{Hz}$, which is insufficient to cover a reasonable portion of the lower hybrid drift wave frequency range. Therefore, we do not include the ion current density $\delta{\bf J}_i$ in this analysis since the high-resolution ${\bf V}_i$ in \Cref{fig:1}b shows negligible fluctuation levels in comparison with ${\bf V}_e$ in \Cref{fig:3}c, indicating that $\delta{\bf J}_i$ should not significantly impact $\delta{\bf J}$. Nevertheless, to test this hypothesis, we down-sampled the electron moments to the high-resolution ion moments sample rate, high-pass-filtered the velocities above $1~\unit{Hz}$, and calculated $\delta{\bf J}$ within the time interval marked in yellow in \Cref{fig:1}. Overlaying each $\delta{\bf J}$ component with its respective for the electron current density $\delta{\bf J}_e$, resulted in a nearly perfect agreement between the curves (not shown), indicating that $\delta{\bf J}_i$ has negligible contribution to the total fluctuating current density within the tested frequency range. In addition, it is expected that at $f>13.3~\unit{Hz}$, $\delta{\bf J}_i$ becomes even less important. Therefore, it is reasonable to approximate $\delta{\bf J} \approx \delta{\bf J}_e$. So, from this point, we drop the subscript $e$ and use only $\delta{\bf J}$ to express the fluctuating electron density current. Below, we present a brief analysis regarding the limitation imposed by the electron moments sample rate to the electric field frequency range.

Figures \ref{fig:3}b-\ref{fig:3}d show an overview of $n_e$, ${\bf V}_e$, and ${\bf E}$ from MMS1. \Cref{fig:3}a shows ${\bf B}$ for context. All vector quantities are in LMN coordinates, with ${\bf V}_e$ (\Cref{fig:3}c) and ${\bf E}$ (\Cref{fig:3}d) decomposed into perpendicular and parallel components with respect to ${\bf B}$. Within the time interval bounded by the dark magenta lines, \Cref{fig:3}c shows intense fluctuations in $V_{e\parallel}$, $V_{eM\perp}$, and $V_{eN\perp}$, with peaks reaching $\sim \pm 1200~\unit{km/s}$. We find that the electrons remain close to frozen in, while ions are unmagnetized, confirming the interpretation of the waves as lower hybrid drift waves \cite{graham2019}. From 08:40:57.5 to 08:40:59.5, we observe four anti-parallel jets that coincide with regions of maximum density in \Cref{fig:3}b, indicating that the spacecraft enters and leaves the electron outflow region of the ongoing magnetopause reconnection as the boundary moves back and forth. In \Cref{fig:3}d, ${\bf E}$ shows intense wave activity, with amplitudes reaching $\sim 50~\unit{mV/m}$, in $E_{M\perp}$ and $E_{N\perp}$, typical of lower hybrid waves at Earth's magnetopause \cite{graham2017,graham2019}.

To evaluate ${\bf J \cdot E}$, we must downsample the electric field to match the current density's sample rate. This means the Nyquist frequency of the high-resolution moments delimits the maximum frequency we are able to access in this analysis, resulting in a maximum frequency of $f_{\rm max}=66~\unit{Hz}$. Therefore, in \Cref{fig:3}e we show the electric field wave power spectrum, in its original sample rate ($8192~\unit{Hz}$), overlaid with $f_{ce}$ (red line), $f_{lh}$ (solid black line), and $f_{cp}$ (blue line). Throughout the interval, the blue line shows $f_{cp} < 1~\unit{Hz}$, which justifies the choice of $f_{\rm min}=1~\unit{Hz}$ as the lower frequency boundary for the electric field oscillations. Most of the intense wave activity in \Cref{fig:3}e occurs from $f_{\rm min}=1~\unit{Hz}$ until the black dotted line marking $f_{\rm max} = 66~\unit{Hz}$. The few remaining amplitude peaks above the black dotted line extend until $f \approx 120~\unit{Hz}$. However, these peaks are sparsely distributed, and the wave power rapidly decreases as the frequency increases. Therefore, even though some amount of wave power is inaccessible to our analysis due to the necessary downsampling on electric field data, most of the lower hybrid wave power is concentrated within $f_{\rm min} \leqslant f \leqslant f_{\rm max}$.

\subsection{Spectral analysis method}
\label{sec:specan}
We are interested in quantifying the energy flow between electron and field \emph{fluctuations}. In such a case, an accurate result must consider both the amplitudes and the phase difference between the fluctuating quantities. Therefore, rather than evaluating ${\bf J \cdot E}$ directly from the time series, we assume harmonic dependence for the current density fluctuations $\delta {\bf J}(t)$ and electric field fluctuations $\delta {\bf E}(t)$:
\begin{align}
  \delta {\bf J}(t) &= {\rm Re}[\delta{\bf \hat J}(\omega)e^{-i\omega t}]\nonumber\\
    \label{eq:1}
  \delta {\bf E}(t) &= {\rm Re}[\delta{\bf \hat E}(\omega)e^{-i\omega t}].
\end{align}
Above, $\delta {\bf \hat J}(\omega)$ is the complex Fourier amplitude of the electron current density, $\delta {\bf \hat E}(\omega)$ is the complex Fourier amplitude of electric field fluctuations, and ``Re'' denotes the real part of the respective quantities.

To evaluate the product of the complex functions in \Cref{eq:1}, we take the time average over a wave period for real $\omega$ \cite{swanson2003}. Then, dropping the ``\,${\bf \hat {}}$\,'' symbol to simplify the notation, we obtain
\begin{equation}
  \label{eq:2}
  \langle \delta{\bf J}(\omega) \cdot \delta{\bf E}(\omega)\rangle = \frac{1}{2}{\rm Re}\left[ \delta{\bf J}^*(\omega) \cdot \delta{\bf E}(\omega)\right],
\end{equation}
where the $*$ denotes the complex conjugate of $\delta {\bf J}(\omega)$, and where $\langle\, \rangle$ represents the time average over a wave period of the enclosed quantities.

In \Cref{eq:2}, the sign and magnitude of $\langle \delta{\bf J}(\omega) \cdot \delta{\bf E}(\omega)\rangle$ are determined by the amplitudes of $\delta {\bf J}^*(\omega)$ and $\delta {\bf E}(\omega)$ and phase differences between both quantities. For instance, if $\delta {\bf J}^*(\omega)$ and $\delta {\bf E}(\omega)$ are $90^{\degree}$ out-of-phase, $\langle \delta{\bf J}(\omega) \cdot \delta{\bf E}(\omega)\rangle = 0$, corresponding to zero energy flow between waves and particles. 

To evaluate \Cref{eq:2}, we apply a complex wavelet transform for frequencies between $f_{\rm min} \leqslant f \leqslant f_{\rm max}$ on both datasets to obtain $\delta {\bf J}^*$ and $\delta {\bf E}$. The result is a time-frequency spectrogram of $\langle\delta{\bf J}(f,t) \cdot \delta{\bf E}(f,t)\rangle$, which describes the period averaged energy flow between electrons and lower hybrid waves measured in $\left[\unit{nW m^{-3}Hz^{-1}}\right]$.

Next, we estimate the total wave period averaged energy flow as a function of time by integrating $\langle \delta{\bf J} \cdot \delta{\bf E}\rangle$ from $f_{\rm min}$ to $f_{\rm max}$:
\begin{equation}
  \label{eq:3}
  \langle \delta{\bf J}(t) \cdot \delta{\bf E}(t)\rangle = \int_{f_{\rm min}}^{f_{\rm max}} \langle\delta{\bf J}(f,t) \cdot \delta {\bf E}(f,t)\rangle\, df,
\end{equation}
where $df$ is obtained by taking the difference between two adjacent components of the frequency vector in $\langle\delta{\bf J}(f,t) \cdot \delta {\bf E}(f,t)\rangle$, which is given in logarithmic scale.

By evaluating \Cref{eq:3}, we obtain a reliable measure of the total energy flow between $f_{\rm min}$ and $f_{\rm max}$, and its sign, versus time $t$ of data within the interval under analysis. With this quantity in hand, we calculate the cumulative sum over time and obtain the net energy flow within a given interval. Another advantage of having $\langle \delta{\bf J} \cdot \delta{\bf E}\rangle$ in time-series format is that it makes it easier to visualize and compare the results in this paper with previous and future non-spectral ${\bf J \cdot E}$ analyses.

In addition, to verify the contribution of the lower hybrid drift waves to the electron transport inside the thin boundary layer, we use the same spectral method to calculate the N component of the anomalous flow associated to the fluctuations, $V_{N\,{\rm anom}}$, which in its vector form is given by
\begin{equation}
  \label{eq:4}
  {\bf V}_{\rm anom}= \frac{\langle\delta n_e \delta {\bf V}_e \rangle}{\langle n_e\rangle},
\end{equation} 
which is closely related to cross-field electron diffusion.

\subsection{Data analysis}
\label{sec:data-analysis}
\begin{figure}[h!]
  \centering  \includegraphics[width=0.98\linewidth]{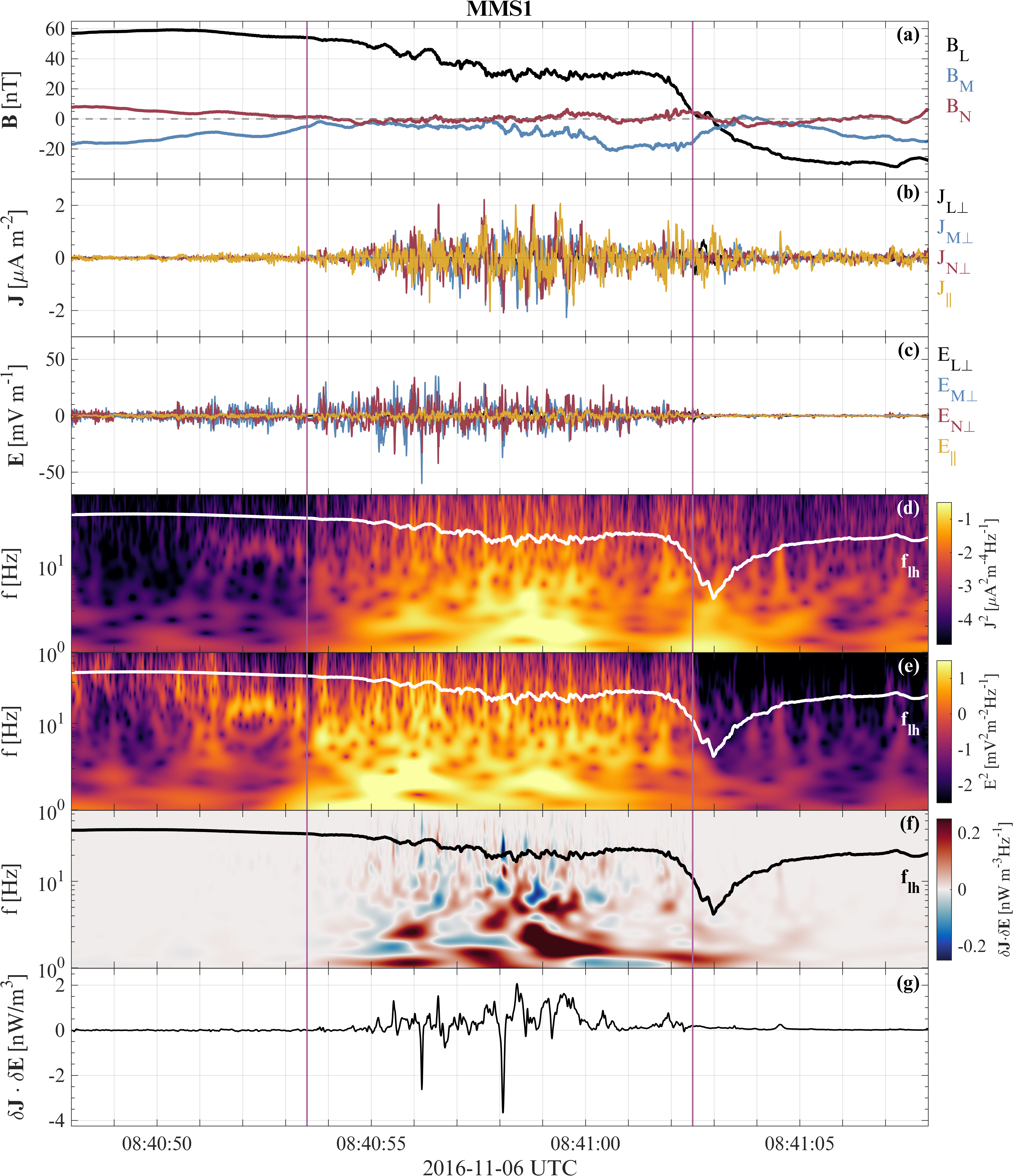}
\caption{Energy flow between lower hybrid waves and electrons as measured by MMS1. (a) ${\bf B}$. (b) ${\bf J}$ from the high-resolution electron moments and band-pass-filtered between $f_{\rm min}\!\leqslant \!f\!\leqslant \!f_{\rm max}$. (c) ${\bf E}$ downsampled and band-pass-filtered to match ${\bf J}$'s sample rate and frequency range. (d) Spectrogram of ${\bf J}$ shown in (b). (e) Spectrogram of ${\bf E}$ shown in (c). (f) Spectrogram of $\langle \delta{\bf J\, \cdot}\, \delta{\bf E}\rangle$, where red represents energy flow from waves to particles, and blue energy flow from particles to waves. (g) Frequency integrated $\langle \delta{\bf J\, \cdot}\, \delta{\bf E}\rangle$. The white line in (d), (e), and (f) shows the lower hybrid frequency for the period. The vertical dark magenta lines mark the same interval as in \Cref{fig:1}.}
\label{fig:4}
\end{figure}

First, we analyze data from MMS1 from 08:40:48 to 08:41:08, as shown in \Cref{fig:4}. The dark magenta lines mark the interval in which the spacecraft travels through the thin boundary layer at the magnetospheric separatrix, highlighted with the same color in \Cref{fig:1}. \Cref{fig:4}a shows ${\bf B}$ for context. \Cref{fig:4}b shows the perpendicular and parallel components of ${\bf J}$. Most of the current density is concentrated between 08:40:55 and 08:41:01, with the highest amplitudes occurring between $\sim$ 08:40:57.5 and $\sim$ 08:40:59.5 in $J_{M\perp}$, $ J_{N\perp}$, and $J_{\parallel}$. In \Cref{fig:4}d, the spectrogram of ${\bf J}$ shows that the highest amplitudes occur at frequencies $< 10~\unit{Hz}$. \Cref{fig:4}c shows the band-pass-filtered perpendicular and parallel components of ${\bf E}$. Comparing ${\bf E}$ in \Cref{fig:4}c with the total electric field wave-form in \Cref{fig:3}d, we notice that the amplitude of $E_{\parallel}$ is significantly smaller, almost vanished, meaning that most of the parallel oscillations have frequencies $>66~\unit{Hz}$ and thus were filtered out during the resampling. Moreover, the downsampling caused a slight amplitude decrease in the perpendicular components of ${\bf E}$. Nevertheless, the overall waveform remained nearly the same as in \Cref{fig:3}d, with the highest amplitudes occurring from $\sim$ 08:40:55 to $\sim$ 08:40:58.2 in $E_{M\perp}$ and $E_{N\perp}$. After 08:40:58.2, ${\bf E}$ becomes more uniform as the amplitude decreases progressively until it reaches a minimum around 08:41:01.5. In \Cref{fig:4}e, the spectrogram of  ${\bf E}$ shows broadband fluctuations with the highest amplitudes occurring for frequencies $\lesssim f_{lh}$, as already discussed in \Cref{sec:method}.

In \Cref{fig:4}f, we show the spectrogram of $\langle\delta{\bf J} \cdot \delta{\bf E}\rangle$ obtained using MMS1 data. The colorbar is centered on $\langle\delta{\bf J} \cdot \delta{\bf E}\rangle=0$, illustrated in off-white color. We notice that most of the figure's background is colored in off-white since nearly all the $\langle\delta{\bf J} \cdot \delta{\bf E}\rangle\neq 0$ is concentrated in the interval within dark magenta lines, where we observe localized red- and blue-hued patches alternating throughout the region. Red-hued patches mark $\langle\delta{\bf J} \cdot \delta{\bf E}\rangle>0$, indicating time and frequency spots where the energy is flowing from waves to electrons. Blue-hued patches mark $\langle\delta{\bf J} \cdot \delta{\bf E}\rangle<0$, and indicate time and frequency spots where the energy is flowing from electrons to waves. The fluctuating energy transfer signatures occur throughout the lower hybrid drift wave frequency range shown in \Cref{fig:4}f, indicating an active energy exchange between particles and waves within the thin boundary layer. We interpret this broadband wave-particle interaction signature collocated with waves in the lower hybrid frequency range, in a plasma mixing layer, as evidence that these waves are generated locally by the lower hybrid drift instability associated with the density gradient inside the thin boundary layer.

In \Cref{fig:4}g, the frequency-integrated energy flow, obtained by evaluating \Cref{eq:3}, shows fast and irregular oscillations around zero for the four spacecraft. We observe large spikes of negative and positive $\langle\delta{\bf J} \cdot \delta{\bf E}\rangle$ until $\sim$ 08:40:57.6, when the signatures with positive energy flow start covering short, but continuous time intervals. Around 08:40:58.3, the energy flow becomes predominantly positive, indicating a clear change from an oscillatory energy exchange to a more dissipative regime. This positive trend is related to an increase in the positive energy flow at all frequencies, which is more visually noticeable at $f<5~\unit{Hz}$, but it also occurs above this frequency range. The average positive energy dissipation suggests the waves are damped as they approach the ion outflow.  After $\sim$ 08:41.00, $\langle\delta{\bf J} \cdot \delta{\bf E}\rangle_{\rm MMS1}$ reduces significantly as both the wave activity and current density diminish and we observe two small positive peaks before MMS1 leaves the thin boundary layer.

It is important to emphasize, however, that the predominantly $\langle\delta{\bf J} \cdot \delta{\bf E}\rangle>0$ at $f<5~\unit{Hz}$ seen in \Cref{fig:4}f is not an established feature of the wave-particle energy exchange at low frequencies. At least not in this event. We analyzed the same plot for the other three spacecraft and they show rather different distribution in frequency of positive and negative $\langle\delta{\bf J} \cdot \delta{\bf E}\rangle$ (not shown). The energy exchange results differ quantitatively for each spacecraft but are qualitatively similar. In \Cref{sec:four-sc-results}, we analyze frequency-integrated results from the four spacecraft.

\subsection{Four spacecraft results}
\label{sec:four-sc-results}
\begin{figure}[h!]
\centering  \includegraphics[width=0.99\linewidth]{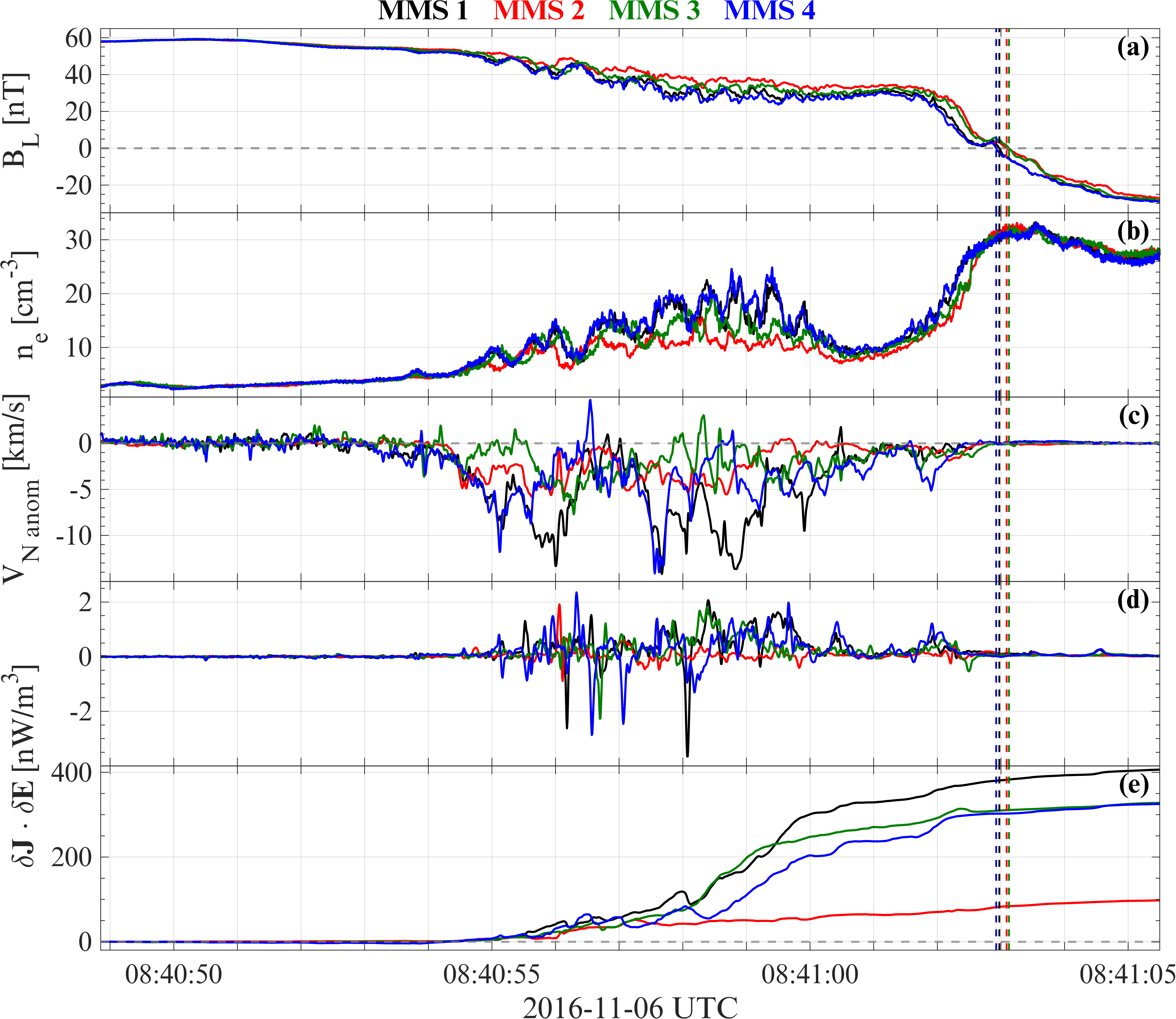}
\caption{Results obtained for the four spacecraft. The lines are color-coded following the legend at the top of the figure. (a) $B_L$, (b) $n_e$, (c) $V_{\rm N\,anom}$, (d) frequency integrated $\langle\delta{\bf J\, \cdot}\,\delta{\bf E}\rangle$, (e) cumulative sum over time of the curves shown in (d). The gray dashed line in (a) marks $B_L=0$, and the vertical dashed lines show the moment each spacecraft crosses $B_L=0$.}
\label{fig:5}
\end{figure}
Now, we compare the estimates of diffusion and energy transfer between the four spacecraft. \Cref{fig:5} is in the time interval highlighted in yellow in \Cref{fig:1}. For context, we show four spacecraft measurements of $B_L$ in \Cref{fig:5}a and of $n_e$ in \Cref{fig:5}b. \Cref{fig:5}c shows the N component of anomalous flow, $V_{N\,\rm anom}$, for the four spacecraft. Like the $\langle\delta{\bf J} \cdot \delta{\bf E}\rangle$ signatures in \Cref{fig:5}d, the anomalous flow occurs only within the time interval corresponding to the thin plasma mixing layer (from 08:40:53.5 to 08:41:02.5). The four spacecraft have highly variable, but predominantly negative $V_{N\,\rm anom}$, which corresponds to magnetosheath electrons diffusing through the magnetopause boundary towards the magnetosphere. MMS1 and MMS4 have the largest peaks of anomalous flow, reaching $\sim 13~\unit{km/s}$, but these peaks are rather localized in time, while MMS2 and MMS3 have lower, but relatively more continuous $V_{N\,\rm anom}$.

In \Cref{fig:5}d, the frequency integrated $\langle\delta{\bf J} \cdot \delta{\bf E}\rangle$ shows that all spacecraft observe highly oscillatory energy exchange. MMS1 and MMS4 have the highest spikes in both directions, followed by MMS3. Comparing with MMS1, $\langle\delta{\bf J} \cdot \delta{\bf E}\rangle_{\rm MMS4}$ has a greater tendency for negative energy flow, meaning MMS4 lies in a more unstable region, with more free energy available to excite the lower hybrid drift instability. However, similarly to what we observed for MMS1, after $\sim$ 08:40:58.4 $\langle\delta{\bf J} \cdot \delta{\bf E}\rangle_{\rm MMS4}$ shifts towards a more dissipative regime, with a predominantly positive energy flow. MMS3 shows similar behavior to MMS1 and MMS4: highly oscillatory and negative spikes until $\sim$ 08:40:58. Then, $\langle\delta{\bf J} \cdot \delta{\bf E}\rangle_{\rm MMS3}$ becomes predominantly positive, shifting to a more dissipative energy flow. On the other hand, MMS2 has a considerably distinct energy exchange signature. It starts oscillating around zero with negligible amplitude until $\sim$ 08:40:56.1, when we observe an intense bipolar signature, with the first peak reaching $\sim + 1.85~\unit{nW/m^3}$, followed by a peak of $\sim - 1.0~\unit{nW/m^3}$. After that, $\langle\delta{\bf J} \cdot \delta{\bf E}\rangle_{\rm MMS2}$ remains oscillating with low amplitudes around zero until MMS2 crosses the magnetosheath. Figures \ref{fig:5}c and \ref{fig:5}d show that electron diffusion toward the magnetosphere occurs regardless of the sign of the electron energy transfer. 

To quantify the net energy flow observed by each spacecraft, we plot in \Cref{fig:5}e the cumulative sum over time of the frequency integrated $\langle\delta{\bf J} \cdot \delta{\bf E}\rangle$. From the start of the figure until $\sim$ 08:40:55, the net energy flow remains very close to zero for the four spacecraft. From 08:40:55 until $\sim$ 08:40:58, MMS1, MMS3 and MMS4 show a small and irregular accumulation. During this interval, MMS1, MMS3 and MMS4 have roughly equivalent net energy flow. MMS4 has the most oscillatory growth, followed by MMS1, as expected from the analysis of \Cref{fig:5}d. MMS3 has a more steady, but slightly slower growth during this interval. Around 08:40:58.3, the net $\langle\delta{\bf J} \cdot \delta{\bf E}\rangle$ growth of MMS1, MMS3 and MMS4 becomes more regular, meaning that the energy flow from that point is predominantly positive, shifting to a more dissipative regime, as discussed in the  \Cref{fig:5}d analysis. After this point, the net $\langle\delta{\bf J} \cdot \delta{\bf E}\rangle$ grows consistently but at different rates for the three spacecraft. The peak net growth rate occurs between $\sim$ 08:40:58.3 and $\sim$ 08:40:59.8 for MMS1 and MMS4, and between $\sim$ 08:40:58.3 and $\sim$ 08:40:59.1 for MMS3. After 08:41:00 the net accumulation rate is significantly reduced, with some localized step-growth. By the end, MMS1 has the highest net $\langle\delta{\bf J} \cdot \delta{\bf E}\rangle$ with $\sim 410~\unit{nW/m^3}$, while both MMS3 and MMS4 have $\sim 325~\unit{nW/m^3}$. As expected, the energy flow accumulation observed by MMS2 tells a completely different story. Around 08:40:56.1, MMS2 has a fast $\langle\delta{\bf J} \cdot \delta{\bf E}\rangle$ net growth, which is related to the bipolar signature shown in \Cref{fig:5}d. From this point, the net energy flow from MMS2 grows at a steady but extremely slow rate, having the lowest net $\langle\delta{\bf J} \cdot \delta{\bf E}\rangle$ by far $\sim 100~\unit{nW/m^3}$ at the end of the interval.

As discussed in \Cref{sec:bound-layer-thickn}, MMS2 is located at the Earthward edge of the plasma mixing layer, where it observes a more magnetospheric-like environment. Therefore, the discrepancy in MMS2 results shown in \Cref{fig:5}, indicates that the energy flow is more efficient inside the plasma mixing layer near the outflow edge bound, where the highest density gradients occur. Nevertheless, all four spacecraft have a positive net energy flow, indicating that electrons gain energy throughout the plasma mixing layer via wave-particle interactions.

\section{Discussion}
\label{sec:discussion}
We observe that lower hybrid drift waves exchange energy with electrons within thin plasma mixing layers, and on average, electrons gain energy from waves. At the same time, we also observe the anomalous flow indicating diffusion of magnetosheath electrons toward the magnetosphere. This diffusion will try to broaden the layer, while ongoing reconnection will try to keep the boundary narrow.

In the mixing region, the density of magnetospheric electrons is very low, and therefore, the magnetosheath electrons have the dominant contribution to the number density. This enhanced population of low-energy electrons diffusing from the magnetosheath into the magnetospheric side leads to the drop in $T_{e\parallel}$ observed between 08:40:55.5 and 08:41:00.0 when MMS is closer to the outflow edge of the mixing layer. However, although much lower than the parallel temperature peaks observed at the mixing layer's edge toward the magnetosphere, the $T_{e\parallel}$ within the densest part of the mixing layer is well above the magnetosheath value. We also find that $T_{e\perp}$ peaks in the mixing region, suggesting that the magnetosheath electrons are heated in the mixing region. However, $T_{e\perp}$ is similar on all spacecraft, despite MMS2 observing significantly lower $\langle\delta{\bf J\,\cdot}\,\delta{\bf E}\rangle$ might suggest the lower hybrid waves do not significantly contribute to the $T_{e\perp}$ increase and other heating processes are operating. 

Previous studies have found evidence that lower hybrid drift waves with a finite parallel phase velocity can interact with electrons via Landau resonance leading to parallel heating \cite{cairns2005,lavorenti2021,ren2022}. In \Cref{fig:3}c, we observe large fluctuations in $V_{e\parallel}$ associated with the lower hybrid drift waves, indicating that the waves have a finite wave-number $k_{||}$ aligned with ${\bf B}$ \cite{graham2019}, meaning that electrons could be heated by the observed lower hybrid drift waves via Landau resonance. 

\begin{figure}[h!]
  \centering  \includegraphics[width=0.99\linewidth]{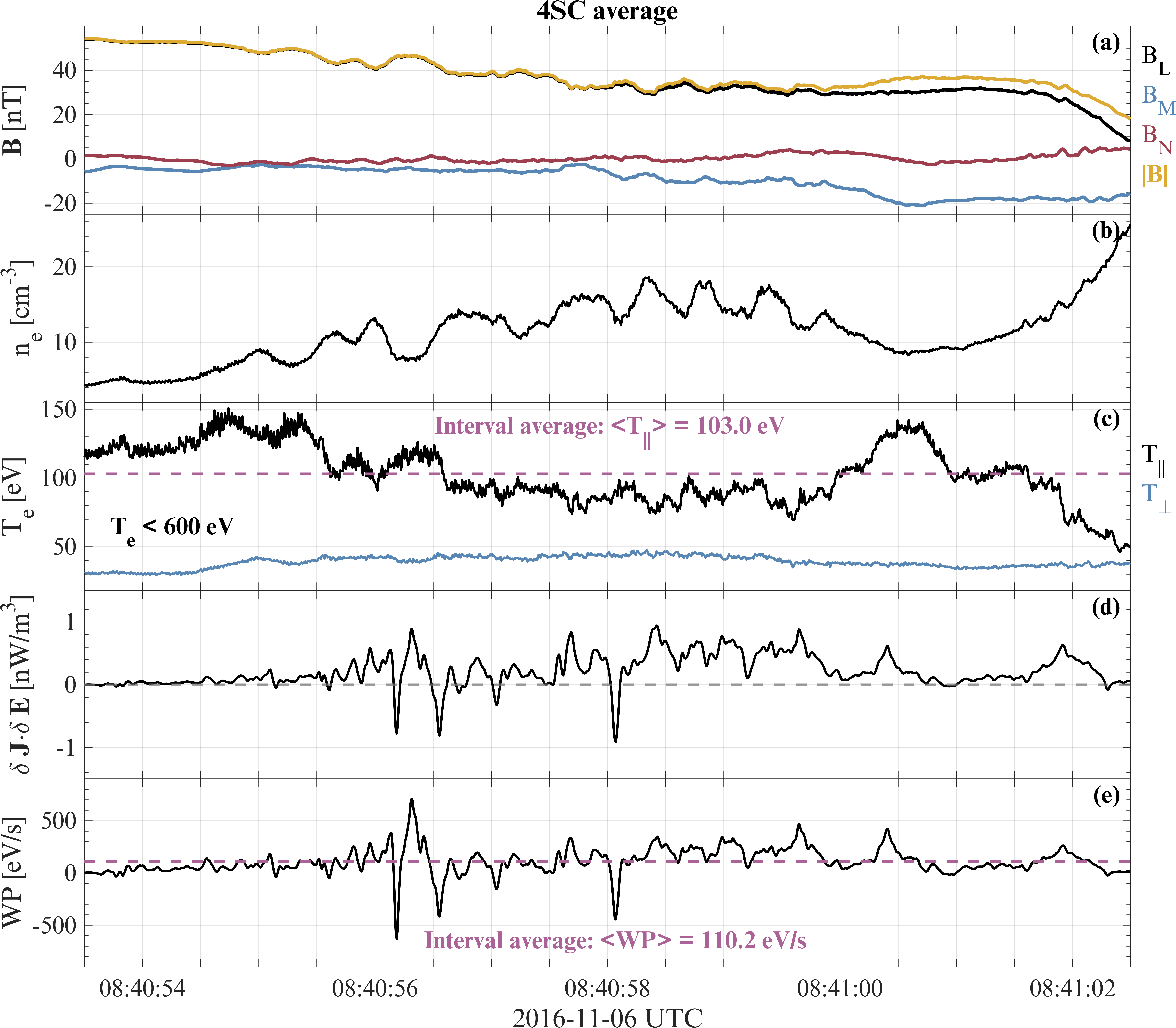}
  \caption{Four-spacecraft averaged quantities within the interval delimited by dark magenta lines in previous figures. (a) ${\bf B}$ and ${\bf |B|}$. (b) $n_e$. (c) $T_{e \parallel}$ and $T_{e \perp}$. (d) $\langle\delta{\bf J\,\cdot}\,\delta{\bf E}\rangle$. (e) Wave power. The dark magenta dashed lines mark the interval averages of the respective quantities.}
  \label{fig:6}
\end{figure}
We now investigate whether the observed wave energy transfer to electrons can account for the observed electron heating. For this analysis, we assume that energy is transferred as heat to the electrons, enabling us to estimate the extent of electron heating due to the waves. To verify if the lower hybrid drift waves can effectively heat the electrons at the plasma mixing boundary, we calculate the four-spacecraft (4sc) average of the wave power ($WP_{4sc}$) for the 4SC-averaged electron number density, as follows
\begin{equation}
  \label{eq:5}
  WP_{4sc} = \frac{\langle\delta{\bf J\, \cdot}\,\delta{\bf E}\rangle_{4sc}}{\langle n_e\rangle_{4sc}}.
\end{equation} 
Since MMS2 has a much lower $\langle\delta{\bf J\, \cdot}\,\delta{\bf E}\rangle$ and measures lower $n_e$ than the other three spacecraft, $WP_{4sc}$ is a conservative estimate for the wave power in this event. In \Cref{fig:6}e, we show $WP_{4sc}$ converted to $\unit{eV/s}$, where the dark magenta dashed line marks the wave power averaged over the interval, $\langle WP\rangle = 110~\unit{eV/s}$. This implies that electrons could gain energy from the waves if they remain in the same region as the waves for a significant amount of time. 

From $\langle WP\rangle$, we can estimate the energy available to heat an electron during the time $\Delta t$ over which it stays in contact with the lower hybrid drift waves. For an estimate of the shortest characteristic time electrons interact with the waves, we consider an electron moving with speed $v_e=2800~\unit{km/s}$, which corresponds to the thermal velocity $v_{Te}=\sqrt{2T_e^{MSH}/m_e}$ of magnetosheath electrons with isotropic temperature $T_e^{MSH}= 22~\unit{eV}$. This would correspond to passing electrons, which are not trapped in the boundary layer. We assume the electron interacts with lower hybrid drift waves with the observed properties while moving parallel to ${\bf B}$ within the plasma mixing layer. Such boundaries can extend over distances of order an ion inertial length $d_i \approx 100~\unit{km}$ \cite{andre2004} along $B_L$, so we define the length of the interaction as $L= d_i\equiv 100~\unit{km}$. Thus, $\Delta t = L_i/v_e = 0.04~\unit{s}$ is the time during which the electron interacts with the lower hybrid drift waves in the plasma mixing boundary. Multiplying $\Delta t$ by $\langle WP\rangle$, we estimate the wave energy available to the electron $E_{we} \sim 4~\unit{eV}$. 

Trapping due to large-scale parallel potentials near the magnetospheric inflow region \cite{egedal2011,egedal2013} could keep electrons in contact with the lower hybrid drift waves for a longer time, thus increasing $\Delta t$. We observe that, within the plasma mixing layer, the electron temperature profile deviates from the scaling laws predicted from the large-scale parallel potential model. These changes in the electron temperature profile are attributed to plasma mixing and transport due to lower hybrid drift waves \cite{graham2017,graham2019,wang2017b,holmes2019,le2018}. In addition, we observe temperature anisotropy with $T_{e\parallel}/T_{e\perp}> 1$ (see \Cref{fig:1}f) and the presence of flat-top velocity distributions (not shown), suggesting that electron heating due to trapping by the large-scale parallel potential persists in the plasma mixing layer. Therefore, we conclude that wave-particle interactions and trapping by large-scale parallel potential are likely active within the plasma mixing layer, meaning that electrons can become trapped in the region where the lower hybrid drift waves occur. 
In the case of electron trapping by large-scale electric fields, the interaction time of electrons with the waves can be estimated as $\Delta t = L_B/V_{in}$, where $L_B \sim 11.5$~km is the thickness of the boundary layer and $V_{in} \sim 40$~km~s$^{-1}$ is the estimated inflow speed, which we use as the speed magnetosheath electrons cross the boundary. We then obtain $\Delta t \sim 0.3$~s, which yields $E_{we} \sim 30~\unit{eV}$. These trapped electrons can gain more energy from the waves than the passing electrons.

To compare with the observed parallel heating, we use the temperature average over the interval $\langle T_{e\parallel}\rangle = 103~\unit{eV}$ (dashed horizontal line in \Cref{fig:6}c) and isotropic magnetosheath temperature $T_e^{MSH}= 22~\unit{eV}$, and obtain a temperature variation of $\Delta T_{e\parallel} = 81~\unit{eV}$. Comparing with the estimated energy gain, $4~\unit{eV} \lesssim E_{we} \lesssim 30~\unit{eV}$, we conclude that heating by lower hybrid waves could be important, but unlikely to account for the increased $T_{e\parallel}$ alone.


The results of this study show that lower hybrid waves transfer energy to electrons and can contribute to electron heating, although this energy transfer is probably not the dominant electron heating process. Rather, the large-scale parallel potentials and the associated electron trapping are likely primarily responsible for the largest $T_{e\parallel}$. Aside from the large-scale trapping potential, processes such as compressional and viscous heating may contribute to the observed electron heating. Further work is needed to identify the specific wave-particle processes responsible for electron heating. Additionally, future studies are needed to model the effectiveness of the trapping by large-scale parallel potential within plasma mixing layers in the presence of strong wave activity. 

\section{Conclusion}
\label{sec:conclusion}
We have used MMS's high-resolution fields and moments data to calculate the energy exchange between electrons and lower hybrid drift waves observed in an electron-scale plasma mixing layer at the edge of the magnetospheric outflow during a slow magnetopause crossing. The observed boundary is thin, and the maximum spacecraft separation along ${\bf \hat N}$ is comparable to the boundary thickness $\sim 6.7\,d_e \sim 11.5~\unit{km}$. In the event analyzed, the four spacecraft approach the magnetopause from the magnetospheric side within the electron edge boundary in a tangent trajectory along the magnetospheric separatrix region, crossing the magnetopause boundary close to but southward of the X-line. We employed complex spectral analysis to directly evaluate and quantify the energy exchange between lower hybrid drift waves and electron current density fluctuations, thus accounting for the phase differences between the fluctuating quantities. Below, we summarize the main findings of this paper:
\begin{enumerate}
  \item We show direct evidence that lower hybrid drift waves exchange energy with electrons within thin plasma mixing layers during magnetopause reconnection. The observed $\langle\delta{\bf J\, \cdot}\,\delta{\bf E}\rangle$ is complex, the energy bounces back and forth between electrons and waves. The frequency-integrated $\langle\delta{\bf J\, \cdot}\,\delta{\bf E}\rangle$ is on average positive, indicating net energy transfer to electrons. On average, an electron gains $100~\unit{eV/s}$ via wave-particle interactions.
  \item The waves produce an anomalous electron flow toward the magnetosphere of $\sim 10 \unit{km/s}$, regardless of the instantaneous sign of the energy exchange. 
  \item The observed energy exchange with lower hybrid waves can contribute to the observed electron heating. However, for wave-particle interactions to significantly heat electrons, the electrons need to remain trapped in the mixing layer. Such trapping can be provided by the large-scale parallel potentials, producing the large temperature anisotropies in the magnetospheric inflow region. 
\end{enumerate}

Overall, we conclude that the observed evolution of electron temperature and density profiles within the thin plasma mixing layer is produced by a combination of electron diffusion across the layer, heating by large-amplitude lower hybrid drift waves, as well as heating and trapping by large-scale parallel potentials.

\section*{Data Availability Statement}
The MMS data used in this study are available at \url{https://lasp.colorado.edu/mms/sdc/public/data/}. The data analysis was performed using the irfu-matlab software package \cite{khotyaintsev2024}.
This study uses scientific-proof color maps from \citeA{crameri2023}, available at \url{https://doi.org/10.5281/zenodo.8409685}.

\acknowledgments
We thank the entire MMS team and instrument PIs for data access and support. This work was supported by the Swedish National Space Agency (SNSA) grant 2021-00135 and by the Knut and Alice Wallenberg foundation (Dnr. 2022.0087).

\bibliography{mms,magrec,books}

\end{document}